\documentclass[12pt]{article}
\usepackage{epsfig}
\newcommand{\lesssim}{\:\mbox{\raisebox{-3pt}{$\stackrel%
{\displaystyle <}{\sim}$}}\:}
\newcommand{\gtrsim}{\:\mbox{\raisebox{-3pt}{$\stackrel%
{\displaystyle >}{\sim}$}}\:}

\begin{document}

\title{\normalsize \hfill UWThPh-2001-6 \\[1cm] \LARGE
4-Neutrino mass schemes\\ and the likelihood of (3+1)-mass spectra}

\author{W. Grimus\footnote{E-mail: grimus@doppler.thp.univie.ac.at} 
\, and T. Schwetz\footnote{E-mail: schwetz@doppler.thp.univie.ac.at}\\
\small Universit\"at Wien, Institut f\"ur Theoretische Physik \\
\small Boltzmanngasse 5, A--1090 Wien, Austria}

\date{20 February 2001}

\maketitle

\begin{abstract}
We examine the (3+1)-class of 4-neutrino mass spectra within a
rigorous statistical analysis based on the 
Bayesian approach to probability.
The data of the Bugey, CDHS and KARMEN experiments are combined
by using a likelihood function.
Our statistical approach allows us to incorporate solar
and atmospheric neutrino
data and also the result of the CHOOZ experiment via
inequalities which involve elements of the neutrino
mixing matrix and are derived from these data. 
For any short-baseline $\Delta m^2$ we calculate a
bound on the LSND transition amplitude $A_{\mu;e}$
and find that, in the $\Delta m^2$--$A_{\mu;e}$ plane,
there is no overlap between the 99\% CL region
allowed by the latest LSND analysis and the region allowed by our
bound on $A_{\mu;e}$ at 95\% CL; there are some small overlap regions
if we take the bound at 99\% CL. Therefore, we conclude
that, with the existing data, the (3+1)-neutrino mass
spectra are not very likely. However, treating the
(2+2)-spectra with our method, we find that they are well compatible
with all data.
\end{abstract}

\newpage

\section{Introduction}

At present, there are three indications in favour of neutrino oscillations
\cite{BP78}, namely the solar $\nu_e$ deficit \cite{sun-exp}, the atmospheric 
$\stackrel{\scriptscriptstyle (-)}{\nu}_{\hskip-3pt \mu}$ 
\cite{SK-atm-98,atm-exp} deficit and the
result of the LSND experiment \cite{LSND,LSND2000} hinting at
$\stackrel{\scriptscriptstyle (-)}{\nu}_{\hskip-3pt \mu} \to
 \stackrel{\scriptscriptstyle (-)}{\nu}_{\hskip-3pt e}$ transitions.
Whereas in the case of the first two indications several experiments agree on
the existence of the effect, the third indication is found only by the LSND
collaboration. Therefore, in many analyses the LSND result is left
out. However, if all three indications in favour of neutrino oscillations are
confirmed, for three mass-squared differences of different orders of magnitude
($10^{-10}\; \mbox{eV}^2\, < \Delta m^2_\mathrm{solar} < 10^{-7} \;
\mbox{eV}^2$ or $\Delta m^2_\mathrm{solar} \sim 10^{-5}\; \mbox{eV}^2$, 
$\Delta m^2_\mathrm{atm} \sim 3 \times 10^{-3}$ eV$^2$,
$\Delta m^2_\mathrm{LSND} \sim 1$ eV$^2$) 
one needs a minimum of four neutrinos, three active ones and a sterile one
\cite{4-early}. In
that case a major revision of our picture of the lepton sector of the
elementary particles would be necessary, with a mixing between the active and
the sterile neutrinos; i.e., 
\begin{equation}\label{mixing}
\nu_{\alpha L} = \sum_{j=1}^4 U_{\alpha j}\, \nu_{jL} 
\quad \mbox{with} \quad \alpha = e, \mu, \tau, s,
\end{equation}
if we stick to the minimum of four neutrinos.
In Eq.~(\ref{mixing}) the left-handed flavour fields are denoted by
$\nu_{\alpha L}$ and the left-handed mass eigenfields by $\nu_{jL}$, and the
$4 \times 4$ neutrino mixing matrix $U$ is assumed to be unitary.

One of the most important issues in the context of 4-neutrino scenarios is the
question of the 4-neutrino mass spectrum \cite{BGG,OY,BGGS}. There are two
different spectral classes with very different properties: the first class
contains four types 
and consists of spectra where three neutrino masses are clustered together,
whereas the fourth mass is separated from the cluster by the mass gap needed
to reproduce the LSND result;\footnote{This class contains the
hierarchical mass spectrum.} the second class has two types where two pairs
of nearly degenerate masses are separated by the LSND gap. These two classes
have been dubbed (3+1) and (2+2)-neutrino mass spectra, respectively
\cite{barger00}. The main difference between these two classes is that, if a
(2+2)-spectrum is realized in nature, the transition into the sterile neutrino
is a solution of either the solar or the atmospheric neutrino problem, 
or the sterile neutrino has to take part in both,
whereas with a (3+1)-spectrum it could be only slightly mixed with the active
neutrinos and mainly provide a description of the LSND result.

It has been argued in the literature
\cite{BGG,OY,BGGS,barger98} that the (3+1)-spectra are strongly disfavoured by
the data, whereas the (2+2)-spectra are the preferred ones, in agreement with
all data showing evidence for neutrino oscillations and also with those where
no such evidence has been found. Recently, in
Refs.~\cite{barger00,carlo,peres} this statement has been challenged because
in the latest LSND analysis the allowed region in the 
$\Delta m^2$--$A_{\mu;e}$ plane, where $A_{\mu;e}$ is the LSND transition
amplitude, has undergone a slight shift towards smaller mass-squared
differences, which makes the (3+1)-spectra somewhat less
disfavoured. Furthermore, in a 2-neutrino analysis of atmospheric neutrino
oscillations the Super-Kamiokande data prefer $\nu_\mu\to\nu_\tau$ conversion
over $\nu_\mu\to\nu_s$ \cite{SK-sobel}. Moreover, there is some
debate also in the solar neutrino problem whether the $\nu_e\to\nu_s$
transition is disfavoured in comparison with other solutions \cite{SK-suzuki}, 
though such a feature seems not to be borne out by a
global analysis of the data \cite{2-sun-GG}. In any case, moving away
from pure 2-neutrino considerations in the solar and atmospheric
neutrino problems, transitions 
into active--sterile superpositions \cite{4-GG,fogli} give viable solutions
to both problems within the (2+2)-spectral schemes, with features which will
be tested in the future \cite{peres}.

The arguments presented in Refs.~\cite{BGG,OY,BGGS}, which disfavour the
(3+1)-mass spectra, are based on exclusion curves from short-baseline (SBL)
experiments, and solar and atmospheric neutrino data
enter into this simplified analysis only through inequalities. 
The advantage of this approach is that its parameters are confined to the
quantities\footnote{Note that in Ref.~\cite{BGGS} the quantities
$c_\alpha = 1-d_\alpha$ are used instead.}
\begin{equation}\label{d}
d_\alpha = |U_{\alpha 4}|^2 \quad (\alpha = e, \mu)
\end{equation}
and the SBL or LSND mass-squared difference $\Delta m^2$. For definiteness we
assume that the mass separated by the LSND gap is $m_4$ and, therefore, 
$\Delta m^2 = |m_4^2 - m_1^2|$. 
It has turned out that the up-down asymmetry of atmospheric multi-GeV
$\mu$-like events measured in the Super-Kamiokande experiment 
\cite{SK-atm-98} is very suitable
to constrain $d_\mu$ \cite{BGGS}, whereas from the solar data it follows that
$d_e$ must be small.
The probability of SBL $\nu_\mu\to\nu_e$
transitions is given by the two-neutrino-like formula \cite{BGG}
\begin{equation}\label{Pmue}
P_{\nu_\mu\to\nu_e} = P_{\bar\nu_\mu\to\bar\nu_e}
= A_{\mu;e} \ \sin^2 \frac{\Delta m^2 L}{4E} \,,
\end{equation}
where
$L$ is the distance between source and detector and $E$ is the
neutrino energy and
\begin{equation}\label{A}
A_{\mu;e} = 4\, d_e d_\mu\,.
\end{equation}
The LSND experiment gives an allowed region in the $\Delta m^2$--$A_{\mu;e}$
plane. 

However, the arguments of  Refs.~\cite{BGG,OY,BGGS} are not based on a
well-defined statistical procedure. Therefore, they remain on a
semi-quantitative level and do not allow to assess a confidence level (CL)
which quantifies the degree at which the (3+1)-spectra are excluded.

In this paper we make a step forward towards such an assessment. The main
points to achieve our goal are the following:
\begin{itemize}
\item The aim is to arrive, for every SBL mass-squared difference 
$\Delta m^2$, at a probability distribution solely in terms of
$d_e$ and $d_\mu$; a suitable method for this purpose is given by 
the likelihood function 
in combination with the Bayesian approach
(see, e.g., Refs.~\cite{cowan,PDG00}).
\item We make full use of the data of the SBL Bugey
\cite{bugey}, CDHS \cite{CDHS} and KARMEN \cite{KARMEN,KARMEN2000}
experiments through the likelihood function.
\item In the spirit of Refs.~\cite{BGG,BGGS}, all information pertaining to
the atmospheric and solar mass-squared differences 
is included via inequalities.
Within our probabilistic framework we are able to treat
inequalities as prior probabilities or with a kind of a maximum likelihood
method. 
\item In this way we treat the inequality following from the atmospheric
up-down asymmetry and, similarly, we include also the result of the CHOOZ
experiment \cite{CHOOZ}; the solar neutrino data allow a simpler treatment in
the context of the Bugey data \cite{bugey}, as
described in Refs.~\cite{BGG,a0}.
\item Eventually, for every given SBL $\Delta m^2$ and any CL $\beta$,
we are able to calculate an upper bound $A^0_\beta (\Delta m^2)$
on the transition amplitude $A_{\mu;e}$, and we can compare such bounds 
$A^0_\beta (\Delta m^2)$ with the 90\% and 99\% CL regions in the 
$\Delta m^2$--$A_{\mu;e}$ plane found by the LSND experiment \cite{LSND2000}. 
\end{itemize}
In the same framework we will also discuss the (2+2)-spectra.

The plan of the paper is as follows. In Section \ref{UDA} we rederive the
atmospheric up-down inequality \cite{BGGS} in a form which is suitable for our
purpose. In Section \ref{B+L} we introduce the likelihood function and the
Bayesian approach, and describe how to incorporate 
inequalities; we apply the methods developed there to the CHOOZ result
and the atmospheric up-down inequality. In Section \ref{plots} 
we explain how we calculate bounds on $A_{\mu;e}$ and discuss
each of the SBL experiments
we use, together with their features which are important in this context.
Details of technical nature are deferred to the appendix.
Our main result, represented as a plot in the $\Delta m^2$--$A_{\mu;e}$ plane,
is also given in this section. In Section \ref{2+2} we consider the
(2+2)-spectra in the framework of our statistical approach and in Section
\ref{concl} we draw our conclusions. 

\section{The atmospheric up-down asymmetry as a constraint 
on short-baseline neutrino oscillations}
\label{UDA}

The most convincing evidence for 
$\stackrel{\scriptscriptstyle (-)}{\nu}_{\hskip-3pt \mu}$ disappearance
in atmospheric neutrino experiments is given by the so-called up-down
asymmetry 
\begin{equation}\label{Aud}
A_\mathrm{ud} = \frac{U-D}{U+D}
\end{equation}
measured by the Super-Kamiokande Collaboration \cite{SK-atm-98}, where $U$ and
$D$ refer to the number of up-going and down-going $\mu$-like multi-GeV events,
respectively. Quoting the number for fully contained events, after
1289 days of operation the result
\begin{equation}\label{Aexp}
A_\mathrm{ud}^\mathrm{exp} =
-0.327 \pm 0.045 \pm 0.004
\end{equation}
was found \cite{chris}. Adding statistical and systematic error in
quadrature, one obtains
\begin{equation}\label{DA}
\Delta A_\mathrm{ud}^\mathrm{exp} \equiv \sigma_A = 0.045 \,.
\end{equation}
Because of the smallness of the systematic error this value is
identical with the statistical error.

Let us now rederive the atmospheric up-down inequality \cite{BGGS}. In
the following we will not indicate antineutrinos but our arguments
will hold for both, neutrinos and antineutrinos. With the assumption
that downward-going atmospheric neutrinos do not oscillate with 
the frequency associated with $\Delta m^2_\mathrm{atm}$ and that
oscillations according to $\Delta m^2$ are averaged out, we obtain
\begin{equation}\label{PD}
P^D_{\nu_\mu\to\nu_\mu} = d_\mu^2 + (1-d_\mu)^2 
\quad \mbox{and} \quad
P^D_{\nu_e\to\nu_\mu} = \frac{1}{2} A_{\mu;e} \,.
\end{equation}
Denoting the number of muon (electron) neutrinos and antineutrinos 
produced in the atmosphere by $n_\mu$ ($n_e$), it follows from
Eq.~(\ref{PD}) that
\begin{equation}\label{D}
D = n_\mu [ d_\mu^2 + (1 - d_\mu)^2 ] + \frac{1}{2} n_e A_{\mu;e} \,.
\end{equation}
For the upward-going neutrinos we have the inequalities \cite{BGGS}
\begin{equation}\label{PU}
P^U_{\nu_\mu\to\nu_\mu} \geq d_\mu^2
\quad \mbox{and} \quad
P^U_{\nu_e\to\nu_\mu} \geq \frac{1}{4} A_{\mu;e} \,.
\end{equation}
We, therefore, have the inequality
\begin{equation}\label{U}
U \geq n_\mu d_\mu^2 + \frac{1}{4} n_e A_{\mu;e} \,.
\end{equation}
Note that for the (3+1)-spectra the amplitude $A_{\mu;e}$ is given by
Eq.~(\ref{A}). 
Since $-A_\mathrm{ud}$ (\ref{Aud}) is a monotonously decreasing
function in $U$, using Eqs.~(\ref{D}) and (\ref{U}) 
we obtain the so-called up-down inequality for
$\mu$-like atmospheric events
\begin{equation}\label{G}
-A_\mathrm{ud} \leq G(d_e,d_\mu) = 
\frac{(1-d_\mu)^2 + d_e d_\mu/r }{(1-d_\mu)^2 + 2d_\mu^2 + 3d_e d_\mu/r} \,.
\end{equation}
In this equation we have defined $r = n_\mu/n_e$. The numerical value
$r \simeq 2.8$ can be read off from Fig.~3 in Ref.~\cite{SK-atm-98} of
the Super-Kamiokande Collaboration.
With similar arguments one can also find an upper bound
\begin{equation}\label{H}
A_\mathrm{ud} \leq H(d_e,d_\mu) = 
\frac{d_\mu(1 - d_\mu) - d_e d_\mu/r }
{1 - d_\mu(1 - d_\mu) + d_e d_\mu/r} \,.
\end{equation}
Hence, the up-down asymmetry is confined to the interval
\begin{equation}\label{range}
-G \leq A_\mathrm{ud} \leq H \,.
\end{equation}

Some remarks are at order. Our inequality (\ref{G}) is a little
different from the analogous inequality in Ref.~\cite{BGGS}, because
in the present case we have not eliminated $d_e$; this is useful
because our aim is to derive a probability distribution in $d_e$ and
$d_\mu$ (see next section). The ratio $r$ has a slight dependence on
the atmospheric zenith angle which is neglected here; however, since
$d_e$ is confined to rather small values \cite{BGG,a0} by the result
of the Bugey experiment \cite{bugey}, the terms containing $r$ are
rather unimportant numerically. The assumptions for deriving
Eqs.~(\ref{G}) and (\ref{H}) are not exactly fulfilled: for down-going
neutrinos with zenith angles around $0^\circ$, oscillations according
to small SBL mass-squared differences $\Delta m^2$ around 0.2 eV$^2$
are not completely averaged out; 
for down-going neutrinos with zenith angles close to $90^\circ$,
oscillations according to $\Delta m^2_\mathrm{atm}$ do occur already, if
the atmospheric mass-squared difference is large. We have checked that
both effects do not change numerically the bound $G$ by more than a
few percent in the worst case. Finally, matter effects, which are
important for up-going neutrinos, do not affect the inequalities
(\ref{PU}) and thus also not the inequality for $U$. Eq.~(\ref{G}) is,
therefore, a
well-established inequality which restricts mainly the allowed range of 
$d_\mu$. Due to Eq.~(\ref{A}), it will be used in the following to constrain
the SBL amplitude $A_{\mu;e}$.

\section{The statistical treatment of inequalities} 
\label{B+L}

As mentioned in the introduction, a suitable method for the purpose 
of deriving a probability distribution in the
variables $d_e$ and $d_\mu$ is given by the likelihood function combined with
the Bayesian approach, which is defined as follows. Suppose one has a
series of measurements $x = (x_1, \ldots , x_n)$ and $r$ parameters
$\theta = (\theta_1, \ldots , \theta_r)$ to be estimated. Then
the Bayesian approach allows to construct a
``posterior'' probability density in the parameter space via (see, e.g.,
Ref.~\cite{cowan,PDG00}) 
\begin{equation}\label{bayes}
p(\theta | x) = \frac{L(x | \theta) \pi(\theta)}{\int d^r \theta' 
L(x | \theta') \pi(\theta')}\,.
\end{equation}
In this expression, $L(x | \theta)$ is the likelihood function and
$\pi(\theta)$ is the prior probability density associated with the
parameters $\theta$, reflecting the state of knowledge of $\theta$
before the measurement. 

\subsection{Inequalities included as priors}
\label{Ba}

Let us suppose now that we consider an observable $Z$ 
whose experimental value is
$z_\mathrm{exp} \pm \sigma_z$ and its \emph{true} value is $z$. 
We assume that the values of measurements of $Z$ are distributed according to
a Gaussian distribution 
\begin{equation}\label{LZ}
L_Z = \frac{1}{\sqrt{2\pi} \sigma_z}
\exp \left[ -\frac{1}{2} 
\left(\frac{z_\mathrm{exp} - z}{\sigma_z}
\right)^2 \right]
\end{equation}
around the true value. Suppose further that from some theoretical
consideration we have the knowledge that the true value $z$ is bounded by 
\begin{equation}\label{priorknow}
a \leq z \leq b \,. 
\end{equation}
We conceive the true value $z$, which is otherwise unknown, as 
a parameter in our scenario and assign to it a prior probability density
\begin{equation}\label{piz}
\pi_Z(z) = \frac{1}{b-a}\, \Theta(b-z)\, \Theta(z-a) \,,
\end{equation}
where $\Theta$ denotes the Heaviside function. Due to lack of further
knowledge, we have assumed a flat prior probability density. Since we are not
interested in a posterior probability density in the parameter $z$ we perform
the integral
\begin{equation}\label{intz}
\int dz L_Z(z) \pi_Z(z) = \ell_Z
\end{equation}
with
\begin{equation}\label{lZ}
\ell_Z = \frac{1}{2(b-a)} \left\{
\mathrm{erf}\,
\left(\frac{b-z_\mathrm{exp}}{\sqrt{2}\,\sigma_z}\right) +
\mathrm{erf}\,
\left(\frac{z_\mathrm{exp}-a}{\sqrt{2}\,\sigma_z}\right)
\right\} \,,
\end{equation}
where we have made use of the error function defined by
\begin{equation}
\mathrm{erf}\,(z) = \frac{2}{\sqrt{\pi}} \int_0^z dt\, e^{-t^2}\,.
\end{equation}
The function (\ref{lZ}) represents then the relevant factor in the posterior
probability density which takes into account the inequality
(\ref{priorknow}). 

To check if we have been lead to a meaningful expression
(\ref{lZ}), we want to discuss the behaviour of this function.
Following from 
$\mathrm{erf}\,(\infty) = -\mathrm{erf}\,(-\infty) = 1$,
we find that for $a \ll z_\mathrm{exp} \ll b$ we have 
$\ell_Z \simeq 1/(b-a)$ to a very good approximation, and for 
$z_\mathrm{exp} \gg b$ or
$z_\mathrm{exp} \ll a$ we have $\ell_A \simeq 0$. In this
discussion, ``much smaller'' or ``much larger'' is defined in units of
$\sigma_z$. If we assume that $\sigma_z$ becomes negligibly small, the
function $\ell_Z$ approaches a step function with $1/(b-a)$ being the height
of the step. The edges of the step
are smoothened out by a finite $\sigma_z$. Thus, we will have a maximal
contribution to the posterior probability density for
$z_\mathrm{exp}$ well inside the interval $[a,b]$, whereas for
$z_\mathrm{exp}$ well outside the interval $[a,b]$ the function $\ell_Z$ is
very close to zero. 
Thus we can be confident that the inequality (\ref{priorknow}) is
reasonably well taken into account by our procedure.

\subsection{Inequalities treated with a maximum likelihood method}
\label{MLa}

Again we depart from Eqs.~(\ref{LZ}) and (\ref{priorknow}). Now we take
into account our lack of knowledge about the true value of $Z$ by
maximizing $L_Z$ as a function of $z \in [a,b]$. 
It is easy to check that one obtains
\begin{eqnarray}
\lefteqn{\max_{z \in [a,b]} L_Z \equiv L^m_Z =} \nonumber \\
&& \frac{1}{\sqrt{2\pi} \sigma_z}
\exp \left\{ -\frac{1}{2} \left[ 
\left( \frac{z_\mathrm{exp}-b}{\sigma_z} \right)^2 \Theta (z_\mathrm{exp}-b) +
\left( \frac{a-z_\mathrm{exp}}{\sigma_z} \right)^2 \Theta (a-z_\mathrm{exp})
\right] \right\} \,. 
\label{LZm}
\end{eqnarray}
In Appendix \ref{relation} we will discuss how this method is related
to the method in the previous section.

\subsection{The treatment of the CHOOZ result and the atmospheric up-down
inequality in our statistical approach} 
\label{CH+ud}

The CHOOZ experiment is a long-baseline $\bar\nu_e$ disappearance
experiment. It measures the survival probability $P_\mathrm{CH}$, for which we
derive the inequality
\begin{equation}\label{CHOOZineq}
P_\mathrm{CH} = 1 - 2\, d_e (1-d_e) - 
2 |U_{e3}|^2 (1 - d_e - |U_{e3}|^2) (1 - \cos \phi_\mathrm{atm})
\leq 1 - 2\, d_e (1-d_e) \,.
\end{equation}
We have used the abbreviation 
$\phi_\mathrm{atm} = \Delta m^2_\mathrm{atm}L/2E$.
We conceive $P_\mathrm{CH}$
in Eq.~(\ref{CHOOZineq}) as the \emph{true} survival probability, as opposed
to the experimental value $P_\mathrm{CH}^\mathrm{exp} = 1.01$ with the error
$\sigma_\mathrm{CH} = 0.039$ \cite{CHOOZ}. Thus we have the range
\begin{equation}\label{rangeCH}
0 \leq P_\mathrm{CH} \leq 1 - 2\, d_e (1-d_e) \,.
\end{equation}
Making the substitutions 
\begin{equation}\label{CHsubst}
a \to 0, \quad b \to 1-2d_e(1-d_e), \quad z \to P_\mathrm{CH}, \quad
z_\mathrm{exp} \to P_\mathrm{CH}^\mathrm{exp}, \quad
\sigma_z \to \sigma_\mathrm{CH} \,,
\end{equation}
the CHOOZ result can be included
in the statistical analysis according to both methods described in the
previous subsections. 

For the atmospheric up-down inequality we substitute
\begin{equation}\label{udsubst}
a \to -G, \quad b \to H, \quad z \to A_\mathrm{ud}, \quad
z_\mathrm{exp} \to A_\mathrm{ud}^\mathrm{exp}, \quad
\sigma_z = \sigma_A \,,
\end{equation}
where $G$, $H$, $A_\mathrm{ud}^\mathrm{exp}$ and $\sigma_A$ are given by
Eqs.~(\ref{G}), (\ref{H}), (\ref{Aexp}) and (\ref{DA}), respectively. 
Notice that, since we have $A_\mathrm{ud}^\mathrm{exp} \ll 0$ 
and $H \geq 0$, the first term in the exponential of the expression
(\ref{LZm}) does not contribute here, and the first term in the expression
(\ref{lZ}) is 1 for all practical purposes.

The substitutions (\ref{CHsubst}) and (\ref{udsubst}) allow us to define
-- according to Eqs.~(\ref{LZ}), (\ref{piz}), (\ref{lZ}), (\ref{LZm}) --
the functions 
$L_\mathrm{CH}$, $\pi_\mathrm{CH}$, $\ell_\mathrm{CH}$, $L^m_\mathrm{CH}$ and
$L_\mathrm{ud}$, $\pi_\mathrm{ud}$, $\ell_\mathrm{ud}$, $L^m_\mathrm{ud}$, 
referring to the CHOOZ experiment and the atmospheric up-down inequality,
respectively. These functions will be used in the following discussions.

\section{A bound on the LSND 
$\stackrel{\scriptscriptstyle (-)}{\nu}_{\hskip-3pt \mu} \to
 \stackrel{\scriptscriptstyle (-)}{\nu}_{\hskip-3pt e}$ 
transition amplitude}
\label{plots}

\subsection{The statistical procedure}
\label{stat}

Let us now describe how to derive our desired probability distribution.
We concentrate on the four parameters
$d_e$, $d_\mu$, $A_\mathrm{ud}$ and $P_\mathrm{CH}$. We want to stress again
that the latter two quantities are conceived as \emph{true} values.
Therefore, they are parameters within our procedure and we will
treat them according to the methods described in Section \ref{B+L}, in
order to arrive at a distribution solely in $d_e$ and $d_\mu$.
The likelihood function is given by
\begin{equation}\label{L}
L = 
L_\mathrm{osc}(d_e,d_\mu) \times
L_\mathrm{ud}(A_\mathrm{ud}) \times L_\mathrm{CH}(P_\mathrm{CH})\,,
\end{equation}
where the first factor $L_\mathrm{osc}$ is the product of the likelihood
functions of the Bugey, CDHS and KARMEN experiments. This likelihood function
will be discussed in the next section and in Appendix \ref{app}.
For the treatment of parameters other than $d_e$ and $d_\mu$, which
appear in the fitting procedure to the SBL experiments, see also Appendix
\ref{app}. 

The physically allowed region
$\mathcal{R}_d$ of $d_e$ and $d_\mu$ is described by the inequalities
\begin{equation}\label{Rd}
\mathcal{R}_d : \quad d_e \geq 0, \; d_\mu \geq 0 
\;\; \mbox{and} \;\; d_e + d_\mu \leq 1 \,.
\end{equation}
In order to incorporate it into our procedure, we define a function
$R(d_e,d_\mu)$ such that
$R(d_e,d_\mu) = 1$ for $(d_e,d_\mu) \in \mathcal{R}_d$ and 0
otherwise. This function has thus the task of a prior which confines
$(d_e,d_\mu)$ to the physically meaningful region.
Adopting the maximum likelihood method of Section \ref{MLa} in order to deal
with the parameters $A_\mathrm{ud}$ and $P_\mathrm{CH}$ and combining this
method with Eq.~(\ref{bayes}), we finally
arrive at the desired probability distribution
\begin{equation}\label{pddm}
p_m(d_e,d_\mu) = \frac{L_\mathrm{osc}(d_e,d_\mu)\, 
L^m_\mathrm{ud}(d_e,d_\mu)\, L^m_\mathrm{CH}(d_e)\,
R(d_e,d_\mu)}{\int dd_e' \int dd_\mu'\,
L_\mathrm{osc}(d_e',d_\mu')\, L^m_\mathrm{ud}(d'_e,d'_\mu)\,
L^m_\mathrm{CH}(d'_e)\, R(d_e',d_\mu')} \,.
\end{equation}
The dependence of $L^m_\mathrm{ud}(d_e,d_\mu)$ and
$L^m_\mathrm{CH}(d_e)$ on $d_e$ and $d_\mu$ comes in through the 
substitutions (\ref{udsubst}) and (\ref{CHsubst}) of the boundaries $a$ and
$b$. 

Discussing now the Bayesian approach described in Section \ref{Ba},
we have the prior probability density
\begin{equation}\label{prior}
\pi(d_e,d_\mu,A_\mathrm{ud},P_\mathrm{CH}) =
\frac{1}{2}\, R(d_e,d_\mu)\, \pi_\mathrm{ud}(d_e,d_\mu,A_\mathrm{ud})\,
\pi_\mathrm{CH}(d_e,P_\mathrm{CH}) \,.
\end{equation}
We stress that
\begin{equation}\label{priorcond}
\int dA_\mathrm{ud} \int dP_\mathrm{CH} \,
\pi(d_e,d_\mu,A_\mathrm{ud},P_\mathrm{CH}) = \frac{1}{2}\, R(d_e,d_\mu)
\end{equation}
is fulfilled. This equation tells us that, after integrating
over the other variables, the prior probability density for $d_e$ and $d_\mu$
is uniform. Since these are the variables whose distribution we want to
calculate, Eq.~(\ref{priorcond}) assures us that with our choice of priors we
have not introduced a bias in the distribution of $d_e$ and $d_\mu$.

According to the Bayesian approach in Section \ref{Ba} we perform the
integrations
\begin{equation}
\int dA_\mathrm{ud} \int dP_\mathrm{CH}\, L_\mathrm{ud}\, 
L_\mathrm{CH}\, \pi\,.
\end{equation}
Finally, we obtain via this method the probability distribution
\begin{equation}\label{pddb}
p_b(d_e,d_\mu) = \frac{L_\mathrm{osc}(d_e,d_\mu)\, 
\ell_\mathrm{ud}(d_e,d_\mu)\, \ell_\mathrm{CH}(d_e)\,
R(d_e,d_\mu)}{\int dd_e' \int dd_\mu'\,
L_\mathrm{osc}(d_e',d_\mu')\, \ell_\mathrm{ud}(d'_e,d'_\mu)\,
\ell_\mathrm{CH}(d'_e)\, R(d_e',d_\mu')} \,.
\end{equation}

It is important to note that for every $\Delta m^2$ we have such a
distribution. The SBL mass-squared difference, which is hidden in
$L_\mathrm{osc}$, is not on the same footing as $d_e$ and $d_\mu$ in
our approach. For every given $\Delta m^2$, we find restrictions on
$d_e$ and $d_\mu$ from experiment\footnote{Note that in the analyses of the
Bugey \cite{bugey} and CDHS \cite{CDHS} experiments $\Delta m^2$ is treated in
the same way.} which allow us finally to obtain a
bound on the transition amplitude $A_{\mu;e}$ for any given CL.
Choosing a CL $\beta$, we find the corresponding bound on $A_{\mu;e}$
by the prescription
\begin{equation}\label{Abound}
\int\limits_{4d_ed_\mu \leq A^0_\beta} \hskip-10pt
dd_e\, dd_\mu\, p_j(d_e,d_\mu) = \beta \quad (j=m,b) \,.
\end{equation}
For instance, if we choose $\beta = 0.99$, with Eq.~(\ref{Abound}) we
can calculate a number $A^0_{0.99}$ such that
$A_{\mu;e} \leq A^0_{0.99}$ at 99\% CL.

The two distributions $p_m(d_e,d_\mu)$ (\ref{pddm}) and $p_b(d_e,d_\mu)$
(\ref{pddb}) are, of course, different functions of the variables. The method
to obtain bounds on $A_{\mu;e}$ as given by Eq.~(\ref{Abound}) is
the same for both distributions. In the following, we will
comment on how much the bounds on $A_{\mu;e}$ calculated with both
distributions differ numerically.

\subsection{A qualitative discussion of the SBL data}
\label{qual}

As mentioned previously, the SBL experiments are treated with
the likelihood function
\begin{equation}\label{LSBL}
L_\mathrm{osc}(d_e,d_\mu) = 
L_\mathrm{Bugey}(d_e) \times L_\mathrm{CDHS}(d_\mu)
\times L_\mathrm{KARMEN}(A_{\mu;e}) \,,
\end{equation}
which enters Eq.~(\ref{L}). Here we want to discuss some features of
$L_\mathrm{osc}$ and how these features influence the bound on
$A_{\mu;e}$. Let us start with the Bugey \cite{bugey} and CDHS
\cite{CDHS} experiments. Both are disappearance experiments and,
therefore, the relevant SBL survival probabilities are given by
\begin{equation}\label{disapp}
P_{\nu_\alpha\to\nu_\alpha} = 
P_{\bar\nu_\alpha\to\bar\nu_\alpha} =
1 - 4\, d_\alpha (1-d_\alpha) \sin^2 \frac{\Delta m^2 L}{4E} \,,
\end{equation}
where $\alpha = e$ refers to the Bugey and $\alpha = \mu$ to the CDHS
experiment. Both experiments have not found evidence for neutrino
oscillations. In the case of the Bugey experiment the survival
amplitude $4 d_e (1-d_e)$ is bounded by 0.1 or smaller values in the
relevant range 
$0.1 \; \mbox{eV}^2 \lesssim \Delta m^2 \lesssim 10 \; \mbox{eV}^2$ at
90\% CL \cite{bugey}. Thus, $d_e$ is either very small or close to
1. Since for solar neutrinos the inequality 
\begin{equation}\label{sun-ineq}
P^\odot_{\nu_e\to\nu_e} \geq d_e^2
\end{equation}
holds \cite{BGG}, $d_e$ must be small\footnote{We want to stress that this is
the only place where solar neutrino data enter our analysis. Moreover,
our inference that $d_e$ must be small is independent of the actual
solution to the solar neutrino problem.} 
and in the fit of the Bugey data
we make the approximation  
$d_e (1-d_e) \simeq d_e$ in the survival probability. In principle,
the inequality (\ref{sun-ineq}) should be included in our analysis
with the help of one of the methods discussed in Sections \ref{Ba} and
\ref{MLa}, but in view of the smallness
of the Bugey survival amplitude this is not necessary. In the case of
the CDHS experiment, for $\Delta m^2 \gtrsim 0.3$ eV$^2$, there is an
analogous feature concerning $d_\mu$ \cite{CDHS}; there, the
selection of the small
$d_\mu$ is guaranteed by the up-down inequality (\ref{G})
\cite{BGGS}. For smaller $\Delta m^2$ the CDHS restriction disappears
and values of $d_\mu$ as large as 0.5 are allowed by the up-down inequality.
Therefore, in $L_\mathrm{CDHS}$ we do not neglect $d_\mu^2$
in the survival amplitude because this would not be justified.

In the KARMEN experiment, $\bar\nu_\mu\to\bar\nu_e$ transitions have not been
observed and the KARMEN exclusion curve \cite{KARMEN,KARMEN2000}
cuts right through the region
allowed by LSND. Therefore, it is important
to take into account the KARMEN result in our analysis. For the
details of our fit to the KARMEN data and the analogous details concerning
the Bugey and CDHS experiments see Appendix \ref{app}.

We include the CHOOZ result in our analysis for $\Delta m^2 \geq 0.05$
eV$^2$, where the bound on $\sin^2 2\theta$ in the CHOOZ plots is a
straight line \cite{CHOOZ}; this guarantees that oscillations with the
SBL $\Delta m^2$ are averaged out and hence Eq.~(\ref{CHOOZineq}) is
valid. The CHOOZ result has an effect on the
$A_{\mu;e}$ exclusion curve mainly for $\Delta m^2$ where the Bugey
bound on $d_e$ is not so strong, which is the case for $\Delta m^2$
around 0.1 eV$^2$ and $\Delta m^2 \gtrsim 5$ eV$^2$. 

\subsection{The bound on $A_{\mu;e}$}
\label{discussion}

\begin{figure}[t]
\begin{center}
\mbox{\epsfig{file=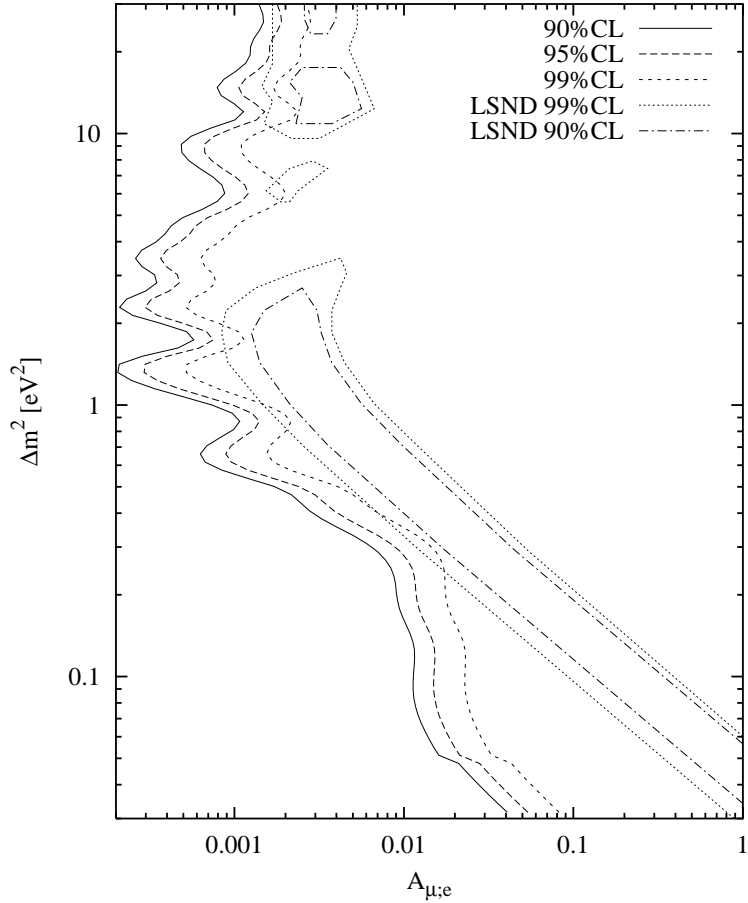,width=0.7\linewidth}}
\end{center}
\caption{Upper bounds on the transition amplitude $A_{\mu;e}$ 
in the case of (3+1)-mass spectra for 90\%, 95\%
and 99\% CL. These bounds have been calculated with the maximum
likelihood approach for the inclusion of the atmospheric up-down
inequality (\ref{G}) and the CHOOZ inequality (\ref{CHOOZineq}) 
as described in Section \ref{MLa}. Also shown are the regions allowed
by the latest LSND results \cite{LSND2000} at 90\% and 99\%
CL. \label{plot1}} 
\end{figure}
Fig.~\ref{plot1} represents the main result of this paper. In this
figure we show the regions in the $\Delta m^2$--$A_{\mu;e}$ plane
allowed by LSND at 90\% and 99\% CL \cite{LSND2000} and our exclusion
curves for $A_{\mu;e}$ with (3+1)-mass spectra at 90\%, 95\% and 99\%
CL. These exclusion curves have been calculated using the maximum
likelihood approach of Section \ref{MLa} for the treatment of the
\emph{true} values $A_\mathrm{ud}$ and $P_\mathrm{CH}$. This method
gives -- for most values of $\Delta m^2$ -- slightly weaker bounds than
the Bayesian approach of Section 
\ref{Ba}; however, the numerical difference between the two methods is
always smaller than 4\%. 
We read off from Fig.~\ref{plot1} that the 99\% CL LSND region has
no overlap with the allowed region to the left of our 95\% CL bound.
This shows that the likelihood of (3+1)-mass spectra is
not very high. Comparing the 99\% CL exclusion curve with the 99\% CL
LSND region and confining ourselves to $\Delta m^2 < 10$ eV$^2$, 
we find overlaps at 
$\Delta m^2 \sim 6, \; 1.7, \; 0.9$ and approximately between 0.25 and
0.4 eV$^2$. This agrees with the findings in Ref.~\cite{carlo}, where
90\% exclusion curves and the bound on $A_{\mu;e}$ derived in Ref.~\cite{BGG}
are compared with the 99\% CL LSND region. 
The result of the inclusion of the CHOOZ result for 
$\Delta m^2 \geq 0.05$ eV$^2$ can be seen as a jump in our exclusion
curves. From this jump the effect of the CHOOZ result can be read off
for small mass-squared differences.
Let us stress once more that in our analysis an exclusion curve for a given
CL is the result of a well-defined statistical procedure and is obtained by
including all available data other than the LSND data; such exclusion curves
have a precise statistical meaning; 
an overlap of the 99\% CL LSND region with the region to the left of
an exclusion curve
occurs only with the exclusion curve at 99\% CL, but not at 95\% CL.

There are claims in the literature that all three indications in favour of
neutrino oscillations are compatible with \emph{three} neutrinos
\cite{3nu}. Such claims have been refuted by a combined fit to the data
\cite{fogli3nu}. In our analysis we arrive at the same conclusion, because our
treatment of the (3+1)-spectra is also applicable in the case of three
neutrinos, where one has (2+1)-mass spectra, since there is a
SBL mass-squared difference $\Delta m^2$ and 
another mass-squared difference much smaller than $\Delta m^2$; assuming that
$m_3$ is the mass separated from the other two by the LSND mass gap, we have
to make the identification $d_\alpha = |U_{\alpha 3}|^2$ ($\alpha =
e,\mu$). The values of $\Delta m^2$ and $A_{\mu;e}$, indicated in the papers
of Ref.~\cite{3nu} as preferred solutions, lie in the region allowed by LSND
at 99\% and thus beyond our 95\% CL exclusion curve in Fig.~\ref{plot1}.

\section{The (2+2)-neutrino mass spectra}
\label{2+2}

If we want to treat the (2+2)-neutrino mass spectra with the method
discussed in the previous two sections, the situation is more
involved. In this case the $\nu_\mu\to\nu_e$ transition amplitude is
given by \cite{BGG}
\begin{equation}\label{A2}
A_{\mu;e} = 4\left| \sum_{j=1,2} U_{ej}U_{\mu j}^* \right|^2 =
4\left| \sum_{j=3,4} U_{ej}U_{\mu j}^* \right|^2
\end{equation}
and cannot be expressed by
\begin{equation}\label{d2}
d_\alpha = \sum_{j=3,4} |U_{\alpha j}|^2
\quad (\alpha = e,\mu) \,,
\end{equation}
as in the case of the (3+1)-spectra (see Eq.~(\ref{A})). 
This suggests to perform an analysis with the five parameters
$d_e$, $d_\mu$, $A_{\mu;e}$, $A_\mathrm{ud}$ and $P_\mathrm{CH}$ in
the case of the (2+2)-spectra.

Let us explore in more detail the relationship between the amplitude
$A_{\mu;e}$ and the parameters $d_e$ and $d_\mu$.
From Eq.~(\ref{A2}), using the
Cauchy--Schwarz inequality, we readily obtain the inequality \cite{BGG,BGG98}
\begin{equation}\label{B}
A_{\mu;e} \leq 4 \min\, [ d_e d_\mu, (1-d_e)(1-d_\mu)] \,.
\end{equation}
This inequality implies that for every $A_{\mu;e}$ the
following region is allowed in the $d_e$--$d_\mu$ plane:
\begin{equation}\label{F}
\mathcal{F}(A_{\mu;e}): \quad \left\{
\begin{array}{rcccl}
\frac{1}{2}(1 - \sqrt{1-A_{\mu;e}}) & \leq & d_e & \leq &
\frac{1}{2}(1 + \sqrt{1-A_{\mu;e}}) \,, \\
A_{\mu;e}/4d_e & \leq & d_\mu & \leq &
1 - A_{\mu;e}/4(1-d_e) \,.
\end{array} \right.
\end{equation}
For every $A_{\mu;e}$ between 0 and 1, this is a region in the unit
square, confined by two hyperbolas. 
Let us label the neutrino masses such that 
$\Delta m^2_{21} = \Delta m^2_\mathrm{atm}$ and
$\Delta m^2_{43} = \Delta m^2_\mathrm{solar}$.
Then, with the
definitions (\ref{d2}), Eqs.~(\ref{D}) and (\ref{U}) hold also for the
(2+2)-mass spectra and we obtain the inequalities 
\begin{equation}\label{G'}
-A_\mathrm{ud} \leq G'(d_e,d_\mu,A_{\mu;e}) = 
\frac{(1-d_\mu)^2 + A_{\mu;e}/4r }{(1-d_\mu)^2 + 2d_\mu^2 + 
3A_{\mu;e}/4r} 
\end{equation}
and
\begin{equation}\label{H'}
A_\mathrm{ud} \leq H'(d_e,d_\mu,A_{\mu;e}) = 
\frac{d_\mu(1 - d_\mu) - A_{\mu;e}/4r }
{1 - d_\mu(1 - d_\mu) + A_{\mu;e}/4r } 
\end{equation}
for the atmospheric up-down asymmetry. Furthermore, the survival
probability for solar neutrinos is bounded by
$P^\odot_{\nu_e\to\nu_e} \geq \frac{1}{2}(1-d_e)^2$. Therefore, 
$d_e$ is close to one\footnote{If $d_e$ is small, then it follows that
$P^\odot_{\nu_e\to\nu_e} \gtrsim 0.5$, which is in disagreement with the
result of the Homestake experiment (see first paper in
Ref.~\cite{sun-exp}); moreover, for small $d_e$ 
we obtain the same bound on $A_{\mu;e}$ as for the disfavoured
(3+1)-mass spectra.} and we make the
approximation $d_e(1-d_e) \simeq 1-d_e$ in the Bugey survival amplitude.
Note that the CHOOZ inequality (\ref{CHOOZineq}) also holds for the
(2+2)-spectra. 

\begin{figure}[t]
\begin{center}
\mbox{\epsfig{file=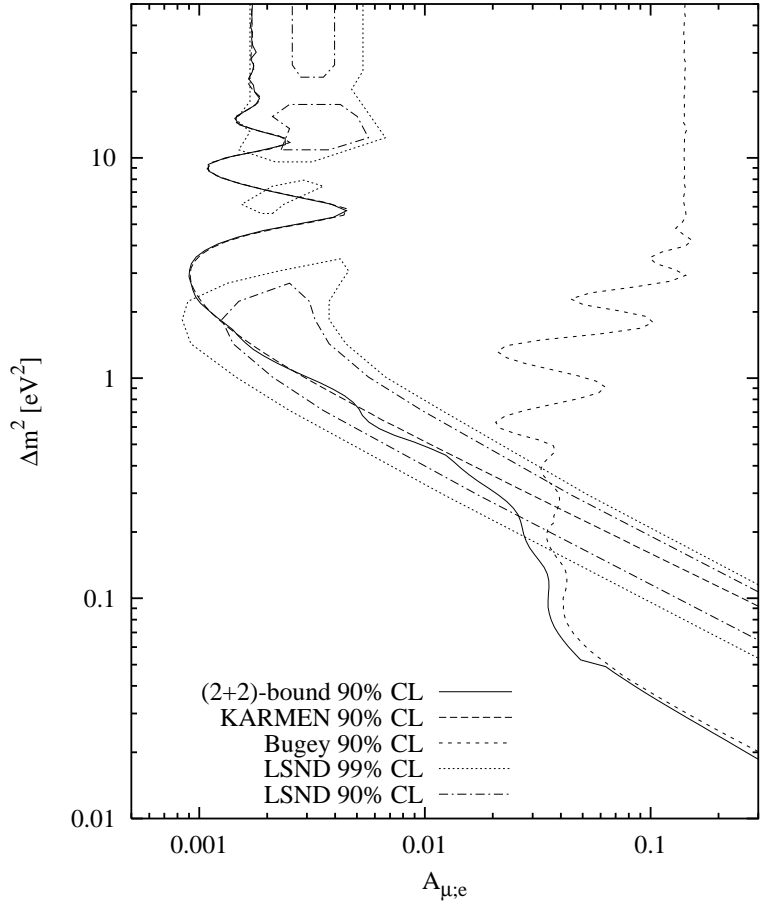,width=0.7\linewidth}}
\end{center}
\caption{The upper bound on the transition amplitude $A_{\mu;e}$ 
in the case of (2+2)-mass spectra for 90\% CL
calculated with the probability distribution (\ref{pm'}). 
Also shown are the 90\% and 99\% CL level regions of LSND of 
Ref.~\cite{LSND2000}, and the
Bugey and KARMEN bounds referring to 90\% CL as given by our
reanalysis. \label{plot2}} 
\end{figure}
In analogy to Eq.~(\ref{Abound}) in the (3+1)-case, our aim is 
to obtain a probability distribution in $A_{\mu;e}$. 
Here we confine ourselves to the maximum likelihood method of Section
\ref{MLa} for the treatment of inequalities. In addition to the
up-down and CHOOZ inequalities we have to take into account
Eq.~(\ref{F}). According to the method of Section \ref{MLa}, we maximize
with respect to $A_\mathrm{ud}$, $P_\mathrm{CH}$, $d_e$ and
$d_\mu$. The maximization with respect to the first two variables
leads to the functions ${L'}^m_\mathrm{ud}$ and $L^m_\mathrm{CH}$,
which are defined in Section \ref{CH+ud}. The prime on
the first function indicates that now it is obtained from
$L^m_\mathrm{ud}$  by the replacement $G \to G'$ and $H \to H'$.
Generalizing the method of Section \ref{MLa}, we define
\begin{equation}
L'(A_{\mu;e}) = \max_{(d_e,d_\mu) \in \mathcal{F}(A_{\mu;e})}
\left[ L'_\mathrm{osc}\, {L'}^m_\mathrm{ud}\, L^m_\mathrm{CH} \right] \,,
\end{equation}
where $L'_\mathrm{osc}$ is given by
Eq.~(\ref{LSBL}), but in view of the appearance of $A_{\mu;e}$ in
$L_\mathrm{KARMEN}$ it is interpreted as a function of three variables.
Finally, we obtain the probability distribution for $A_{\mu;e}$ given by
\begin{equation}\label{pm'}
p'_m(A_{\mu;e}) = 
\frac{ L'(A_{\mu;e}) \Theta (A_{\mu;e})\, \Theta (1-A_{\mu;e})}%
{\int dA'_{\mu;e} \, L'(A'_{\mu;e})\, 
\Theta (A'_{\mu;e})\, \Theta (1-A'_{\mu;e})} \,.
\end{equation}

In Fig.~\ref{plot2} the 90\% CL bound on $A_{\mu;e}$ calculated with
the distribution (\ref{pm'}) is plotted. In this figure the
90\% and 99\% CL level regions of LSND are also depicted. The region to the
left of the bound has an overlap area with the LSND region of 90\%
CL. This shows that the (2+2)-spectra are well compatible with all
data. Also shown in Fig.~\ref{plot2} are the KARMEN and Bugey
exclusion curves. As expected, for large $\Delta m^2$ the solid line
of our exclusion curve follows rather well the KARMEN exclusion curve,
whereas for small $\Delta m^2$ it tends to follow the Bugey curve. Due
to the inclusion of the CHOOZ result above 0.05 eV$^2$, the solid line
is more restrictive there than the Bugey curve, whereas below 0.05 eV$^2$
both curves are nearly identical. We have also investigated the
exclusion curve of the (2+2)-spectra within the Bayesian approach to
inequalities of Section \ref{Ba}. In this approach we have more
freedom to make ``reasonable'' choices for priors than in the
(3+1)-case, and, numerically, the $A_{\mu;e}$ bounds tend to be a
little more restrictive than the solid line in Fig.~\ref{plot2}.

\section{Conclusions}
\label{concl}

In this paper we have examined the (3+1) and (2+2)-classes of the 4-neutrino
mass spectra within a rigorous statistical analysis. Since we do not have
sufficient information concerning the final LSND data \cite{LSND2000}, we have
chosen the approach to analyse all other available data and compare our result
with the LSND result in a $\Delta m^2$--$A_{\mu;e}$ plot. 
This approach suggests
to extract that information from the solar and atmospheric data which is most
relevant for SBL oscillations with respect to $\Delta m^2$. 
This extraction is most
appropriately done in the form of inequalities involving elements of the
neutrino mixing matrix \cite{BGG,OY,BGGS}. Similarly, we have extracted the
``SBL information'' contained in the CHOOZ result. On the other hand,
concerning the SBL experiments, we have fully included the data of the
Bugey, CDHS and KARMEN experiments.

Since our aim has been to incorporate in the statistical analysis the
atmospheric up-down inequality, Eq.~(\ref{G}) or (\ref{G'}), and the
CHOOZ inequality (\ref{CHOOZineq}), 
we have made use of the likelihood function in combination
with the Bayesian approach to probability which allows us to derive
probability distributions of the parameters which are to be
estimated. In this context we have presented two possibilities for
including inequalities involving parameters: one way is to treat them
in the form of prior probability densities for which a ``reasonable
choice'' has to be made; another way is a kind of
maximum likelihood treatment. Numerically, we have compared both
methods for exclusion curves in the case of the (3+1)-spectra and
found that the difference is negligible.

With the method described in the paper we have obtained for every $\Delta m^2$
a probability distribution for the SBL transition amplitude $A_{\mu;e}$, from
which we could derive bounds as a function of $\Delta m^2$ on this amplitude
for any CL. The results are shown in Figs.~\ref{plot1} and \ref{plot2} for the
(3+1) and (2+2)-neutrino mass spectra, respectively. In the latter case our
90\% CL exclusion curve is close to the KARMEN exclusion curve down to 
$\Delta m^2 \sim 0.5$ eV$^2$; there it turns off and starts to come
close to the exclusion curve given by the Bugey data. 
Thus, for the (2+2)-spectra
our method reproduces more or less what one obtains by naively comparing the
KARMEN and Bugey exclusion curves with the region allowed by LSND. Therefore,
these spectra are well compatible with all the data. On the other hand, in the
case of the (3+1)-spectra, our 95\% CL bound has no overlap with the LSND
region of 99\% CL. We, therefore, come to the conclusion that this spectral
class is rather unlikely, even with the recent change in the LSND region.

Thus we strengthen with the method presented here the claims made in
Refs.~\cite{BGG,OY,BGGS}. Should the LSND result be confirmed by the MiniBooNE
Collaboration \cite{MiniBooNE}, then, in a 4-neutrino scheme,
the sterile neutrino should make its
appearance either in the solar or atmospheric neutrinos, or both.

\section*{Acknowledgements}
We thank C. Giunti and S.M. Bilenky for many useful discussions. We
are very grateful to F. Dydak and J. Wotschack for valuable
information about the CDHS experiment and to C. Walter for providing
us with the latest results of the Super-Kamiokande
up-down asymmetry. T.S. is supported by the Austrian Academy of Sciences.

\begin{appendix}

\section{The relationship between the Bayesian and the maximum likelihood
methods for the treatment of inequalities}
\label{relation}

Here we want to elucidate the relationship between the methods for the
inclusion of inequalities introduced in Sections \ref{Ba} and \ref{MLa}.
In the Bayesian approach to this problem we consider in general
integrals of the type
\begin{equation}\label{int}
I[\pi] = \int_a^b dz L_Z(z) \pi(z) \,,
\end{equation}
where $L_Z$ is given by Eq.~(\ref{LZ}). A prior $\pi$ is a positive
and piecewise continuous function such that $\int_a^b dz\, \pi(z) = 1$. In
Section \ref{Ba} we have used the flat prior (\ref{piz}). If we
denote the set of all prior probability densities on the interval
$[a,b]$ by $\mathcal{P}$, then the following proposition holds.\\[2mm]
\textbf{Proposition:} For the integral in Eq.~(\ref{int}) one
has the upper limit
\begin{equation}\label{theorem}
\max_{\pi \in \mathcal{P}} I[\pi] = L_Z^m \,,
\end{equation}
where $L_Z^m$ is given by Eq.~(\ref{LZm}). \\[1mm]
\textbf{Proof:} The proof is very straightforward. Let us first
assume that $z_\mathrm{exp} > b$. Then we have
$$
I[\pi] < L_Z(b) \int_a^b dz\, \pi(z) = L_Z(b) \,.
$$
Thus we have found an upper bound on $I$. If we can find a sequence of
$\pi_k$ of prior probability densities such that
$$
\lim_{k\to\infty} I[\pi_k] =  L_Z(b) \,,
$$
then the proposition is proven for the case $z_\mathrm{exp} > b$. Such
a sequence is easy to give: any sequence $\pi_k$, for which 
$\pi_k(z) = 0$ for $z \leq b-1/k$ holds, fulfills our purpose.
For $z_\mathrm{exp} < a$ we proceed analogously. If
$z_\mathrm{exp} \in [a,b]$, the upper bound on $I$ is
$L_Z(z_\mathrm{exp})$. The sequence $\pi_k$ is constructed accordingly
with the idea given 
before. Thus for the three cases $z_\mathrm{exp} > b$,
$z_\mathrm{exp} < a$ and $z_\mathrm{exp} \in [a,b]$ 
we have the upper limits $L_Z(b)$, $L_Z(a)$ and $L_Z(z_\mathrm{exp})$,
respectively. In summary, we just have obtained the expression
(\ref{LZm}). Q.E.D.

Thus the expression $L_Z^m$ is obtained from the Bayesian approach as
the maximum over all possible priors. Sloppily speaking, the prior in
$I$, which gives $L_Z^m$, is a delta function ($\delta (z-b)$, 
$\delta (z-a)$ or $\delta (z-z_\mathrm{exp})$); however, in order to
choose the correct delta function one has to know the experimental
value $z_\mathrm{exp}$ and, therefore, such a function does not
deserve the name ``prior'' anymore.
Though $\ell_Z < L_Z^m$ holds, it is not obvious which method gives the
stronger constraint in an actual situation, 
because one has additional factors in the probability
distribution one aims at, coming from additional data, and one has to
normalize the combined distribution (see Eq.~(\ref{bayes})).
Moreover, the bounds $a$ and $b$ are in general functions of the parameters
whose distribution we want to know. In the concrete situation discussed in
Section \ref{plots} the maximum likelihood method represented by $L_Z^m$ gives
a slightly weaker restriction on the transition amplitude $A_{\mu;e}$
for most $\Delta m^2$.

\section{The analyses of the Bugey, CDHS and KARMEN experiments}
\label{app}

In this appendix we describe how the data of the
SBL experiments Bugey, CDHS and
KARMEN is included in our analysis. We use as much
information as can be recovered from the publications of
the experimental groups to perform a fit to the data.
The fact that we can reproduce the published 90\% CL bounds in the
case of 2-neutrino oscillations to a good accuracy
inspires confidence in our analysis.

\subsection{Bugey}
The Bugey experiment \cite{bugey} searches for $\bar\nu_e$ disappearance
at the three distances 15 m, 40 m and 95 m away from a nuclear reactor.
The electron antineutrinos are detected through the reaction
$\bar\nu_e + p \to e^+ + n$.
As input data for our analysis we use  Fig.~17 of Ref.~\cite{bugey},
where the ratios of the observed events to the number of expected events
in case of no oscillations are shown for the three positions
in bins of positron kinetic energy in the range 
1 MeV $\le E_{e^+} \le$ 6 MeV.

For the analysis we follow Eq.~(9) of Ref.~\cite{bugey} and
use the $\chi^2$-function
\begin{equation}
\chi^2 = \sum_j \left\{
\sum_{i=1}^{N_j} \frac{ \left[ (A a_j + b(E_{ji}-E_0))R_{ji}^{\mathrm{theo}}
- R_{ji}^{\mathrm{exp}} \right]^2 }{\sigma_{ji}^2} +
\frac{(a_j - 1)^2}{\sigma_a^2} \right\}
+ \frac{(A-1)^2}{\sigma_A^2} + \frac{b^2}{\sigma_b^2} \,.
\label{chi2bugey}
\end{equation}
Here $j=15,40,95$ labels the three positions, $i$ the positron energy bins and
$N_{15}=N_{40}=25,\:N_{95}=10$ are the numbers of bins at each position.
For $E_{ji}$ we take the mean positron energy in bin $ji$.
$R_{ji}^{\mathrm{exp}}$ is the ratio of measured to expected events in
each bin with its statistical error $\sigma_{ji}$,
both read off from Fig.~17 of Ref.~\cite{bugey}.
$R_{ji}^{\mathrm{theo}}$ is the theoretical prediction for this ratio
in the case of oscillations, depending on the oscillation parameters,
and we set $R_{ji}^{\mathrm{theo}} =
\langle P_{\bar\nu_e \to \bar\nu_e} \rangle_{ji}$,
where $P_{\bar\nu_e \to \bar\nu_e}$ is given in Eq.~(\ref{disapp}).
Various systematic uncertainties are taken into account by
minimizing the $\chi^2$, Eq.~(\ref{chi2bugey}), with respect to the
five parameters $A$, $a_j$ and $b$ for given oscillation parameters.
In Ref.~\cite{bugey} the values
$\sigma_A=4.796\%, \, \sigma_a=1.414\%, \, \sigma_b=0.02$ MeV$^{-1}$ and
$E_0=1$ MeV are given.

To perform the averaging of the survival probability
we estimate the uncertainty in the flight length of the neutrinos
because of the size of the production region and the detector to 3 m
and we assume that the flux varies with the distance as $L^{-2}$.
In the relevant energy range antineutrino and positron energy are related by
$E_\nu = E_{e^+} + 1.8$ MeV to a very good approximation.
For the purpose of averaging the survival probability
over the energy range in one bin it is a good approximation to take
neutrino flux, detection cross section and efficiency
as constant with energy. The reason for this is that the
energy bins are relatively small and only ratios of observed to expected
events in each bin are considered. Furthermore, 
we assume a Gaussian resolution
function for the positron energy measurement with variance 0.4 MeV.

The likelihood function in Eq.~(\ref{LSBL}) which
contains the information of the Bugey experiment
is obtained from the $\chi^2$ of Eq.~(\ref{chi2bugey})
by \cite{PDG00}
\begin{equation}
L_{\mathrm{Bugey}}(d_e) \propto \exp\left(-\frac{1}{2}\chi^2 \right)\,.
\end{equation}
The 90\% CL bound on $d_e$ obtained from this likelihood function alone
can be compared to the Bugey exclusion curve in the 2-neutrino case
with the identification $\sin^22\theta_{\mathrm{Bugey}} \simeq 4d_e$.
The curve obtained by our analysis is shown in
Fig.~\ref{plot2} and compares well with the originally published one
\cite{bugey}.

In Eq.~(\ref{disapp}) contributions of oscillations because of
$\Delta m^2_{\mathrm{atm}}$ to the SBL disappearance amplitude are
neglected. This approximation may not be exactly fulfilled in the Bugey
experiment for small $\Delta m^2_{\mathrm{SBL}}$ and
large $\Delta m^2_{\mathrm{atm}}$. We have calculated
the 90\% CL bound from Bugey by taking into account also oscillations with
$\Delta m^2_{\mathrm{atm}}=6\times 10^{-3}$ eV$^2$, which is the 99\% CL
upper bound on $\Delta m^2_{\mathrm{atm}}$ \cite{SK-sobel}, and find that for
$\Delta m^2_{\mathrm{SBL}} > 0.04$ eV$^2$ the effect is smaller than 6\%.

\subsection{CDHS}
\label{subCDHS}

\begin{figure}[t]
\begin{center}
\mbox{\epsfig{file=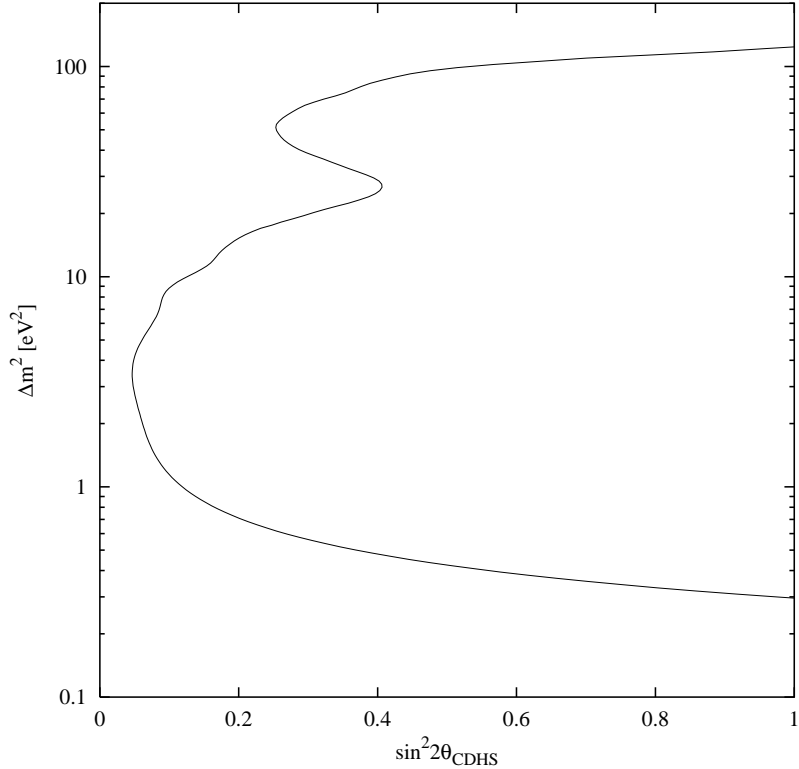,width=0.7\linewidth}}
\end{center}
\caption{The 90\% CL bound on $\sin^2 2\theta_\mathrm{CDHS}$ 
given by our reanalysis of the CDHS experiment as described in Section
\ref{subCDHS}. \label{plot3}} 
\end{figure}
The CDHS experiment \cite{CDHS} searches for $\nu_\mu$ disappearance
by comparing the number of events in the so-called back and front
detectors at the distances $L_{\mathrm{back}} = 885$ m and
$L_{\mathrm{front}} = 130$ m, respectively, 
from the neutrino source. The neutrinos
are detected via muons produced in charged-current interactions.
The data is given in form of the double ratios
\begin{equation}\label{Rcorr}
R_{\mathrm{corr}} = \frac{N_{\mathrm{back}}/N_{\mathrm{front}} }
{N_{\mathrm{back}}^{\mathrm{MC}}/N_{\mathrm{front}}^{\mathrm{MC}}} \,,
\end{equation}
where $N_{\mathrm{back}}\:(N_{\mathrm{front}})$ is the number of observed
events in the back (front) detector and
$N_{\mathrm{back}}^{\mathrm{MC}}$ and $N_{\mathrm{front}}^{\mathrm{MC}}$
are the corresponding quantities expected for no oscillations calculated
by Monte Carlo.

The ratios (\ref{Rcorr}) are given in 15 bins of
``projected range in iron''. This is the distance traveled by the muon
in the detector (consisting of iron) projected onto the detector axis,
which has an angle of $22^\circ$ relative to the neutrino beam axis.
We calculate the range in iron $r(E_\mu)$ of a muon with energy $E_\mu$ by
integrating Eq.~(23.1) of Ref.~\cite{PDG00}. The muon energy intervals
$[E_\mu^{(i1)}, E_\mu^{(i2)}]$ corresponding to the intervals of
projected range ($r_{\mathrm{proj}}$) for bin $i$
given in Table 1 of Ref.~\cite{CDHS} are obtained by applying the relation
$r_{\mathrm{proj}}(E_\mu) \simeq r(E_\mu)\cos\vartheta\cos 22^\circ$,
where $\vartheta \simeq 20^\circ$ is the average scattering angle of
the muons \cite{Wotschack}.

We estimate the number of events in bin $i$ and at position
$p = \:$back or front using
\begin{equation}\label{NCDHS}
N_{ip} \propto M_p \int_{E_\mu^{(i1)}}^{E_\mu^{(i2)}} dE_\mu
\int_{E_\mu}^\infty dE_\nu \int_{L_p-\Delta L_p/2}^{L_p+\Delta L_p/2} dL
L^{-2} P_{\nu_\mu \to \nu_\mu}(L/E_\nu) \Phi(E_\nu)
\frac{d\sigma_{\mathrm{DIS}}(E_\nu,E_\mu)}{dE_\mu} \,.
\end{equation}
Here $M_p$ is the detector mass at position $p$.
Taking into account the size of the decay
tunnel (52 m) and the length relevant for detection in the
back/front detector (72 m/22 m) \cite{CDHS} we have
$\Delta L_{\mathrm{back}}=52+72$ m and $\Delta L_{\mathrm{front}}=52+22$ m.
The disappearance probability $P_{\nu_\mu \to \nu_\mu}(L/E_\nu)$
depending on the oscillation parameters is given in Eq.~(\ref{disapp})
and $\sigma_{\mathrm{DIS}}$ is the cross section for the deep 
inelastic scattering process $\nu_\mu + N \to \mu + X$ 
(see, e.g., Ref.~\cite{bilenky}). The neutrino flux $\Phi(E_\nu)$ in the
relevant neutrino energy range $E_\nu \gtrsim 0.7$ GeV is proportional to
$\exp(-E_\nu/1\, \mathrm{GeV})$ \cite{CDHS,Wotschack}.
Finally, we obtain for the theoretical prediction for the 
double ratios (\ref{Rcorr}) in bin $i$ 
\begin{equation}
R_{\mathrm{theo}}^i=\frac{N_{i\mathrm{back}}}{N_{i\mathrm{front}}}
\left(\frac{L_{\mathrm{back}}}{L_{\mathrm{front}}}\right)^2
\frac{M_{\mathrm{front}}}{M_{\mathrm{back}}}\,,
\end{equation}
and we define the CDHS likelihood function by
\begin{equation}
L_{\mathrm{CDHS}}(d_\mu)\propto \exp\left[ -\frac{1}{2}
\sum_{ij} (R_{\mathrm{corr}}^i - R_{\mathrm{theo}}^i) (S^{-1})_{ij}
(R_{\mathrm{corr}}^j - R_{\mathrm{theo}}^j) \right]\,.
\end{equation}
Assuming total correlation between any two bins, the
covariance matrix is given by
$S_{ij} = \delta_{ij} \sigma_i^2 + \sigma^2_{\mathrm{syst}}$.
$R_{\mathrm{corr}}^i$ and its statistical error $\sigma_i$
are read off from Table 1 of Ref.~\cite{CDHS}. 
The overall systematic error in the ratio of event
rates in the two detectors was estimated to
$\sigma_{\mathrm{syst}}=2.5$\% in Ref.~\cite{CDHS}.  

The 90\% CL bound on $\sin^22\theta_{\mathrm{CDHS}}=4d_\mu(1-d_\mu)$
obtained by our analysis is shown in Fig.~\ref{plot3}. It is very
similar to the bound published by the CDHS collaboration \cite{CDHS}.
There are minor differences for small mass squared differences, which
could have some effect for our bound on $A_{\mu;e}$ in the region 
0.2 eV$^2 \lesssim \Delta m^2 \lesssim 0.4$ eV$^2$: our bound disappears at
$\Delta m^2 \simeq 0.3$ eV$^2$, whereas the CDHS bound extends down to
approximately 0.24 eV$^2$. 

\subsection{KARMEN}

The latest data of the $\bar\nu_\mu \to \bar\nu_e$ oscillation
search in the KARMEN experiment is presented in Ref.~\cite{KARMEN2000}.
Analysing the data taken from February 1997 to March 2000 they find
a total of 11 candidate events, in good agreement with the expected
number of background events for no oscillations of $12.3\pm 0.6$.
For our analysis we use the data resulting from the detection
process $\bar\nu_e + p \to e^+ + n$. The positron spectrum
$S(E_{e^+})$ expected for $A_{\mu;e}\equiv\sin^22\theta_{\mathrm{KARMEN}} = 1$
and $\Delta m^2 = 100$ eV$^2$ we take from Fig.~2(a) of Ref.~\cite{KARMEN}.
To estimate the number of events in a positron
energy interval $[E_1,E_2]$ resulting from neutrino oscillations
we follow Eq.~(B1) of Ref.~\cite{FLS97}:
\begin{equation}\label{Nosc}
N^{\mathrm{osc}} = N \int_{E_1}^{E_2} dE_{e^+} \, S(E_{e^+})
\int_{L_1}^{L_2} dL \, L^{-2} P_{\bar\nu_\mu\to\bar\nu_e}(L/E_\nu)\,,
\end{equation}
where the oscillation probability $P_{\bar\nu_\mu\to\bar\nu_e}$
is given in Eq.~(\ref{Pmue}).
For $L_{1,2}$ we take $17.5\mp 1.75$ m \cite{FLS97} and antineutrino energy
and positron kinetic energy are related by $E_\nu = E_{e^+} + 1.8$ MeV.
The normalization factor $N$ is fixed by requiring that for the total
positron energy range $E_1=16$ MeV, $E_2=52$ MeV, full mixing
($\sin^22\theta_{\mathrm{KARMEN}} = 1$) and $\Delta m^2 \ge 100$ eV$^2$
the number of events resulting from oscillations is 2442 (see Table 1 of
Ref.~\cite{KARMEN2000}).
                  
From Fig.~2(b) of Ref.~\cite{KARMEN2000} we read off for each of the 9
positron energy bins the number of observed events $N_i^{\mathrm{obs}}$
and the number of background events expected for no oscillations $B_i$. Then
we construct the likelihood function by using the Poisson distribution:
\begin{equation}\label{LKARMEN}
L_{\mathrm{KARMEN}}(A_{\mu;e})=\prod_{i=1}^9 \frac{1}{N_i^{\mathrm{obs}}!}
(N_i^{\mathrm{osc}}+B_i)^{N_i^{\mathrm{obs}}} e^{-(N_i^{\mathrm{osc}}+B_i)}\,.
\end{equation}
Here $N_i^{\mathrm{osc}}$ is calculated from Eq.~(\ref{Nosc}) by choosing
$E_1$ and $E_2$ according to the bin $i$.
The 90\% CL bound on $A_{\mu;e}$ for a given $\Delta m^2$
obtained from the probability distribution implied by Eq.~(\ref{LKARMEN})
in the Bayesian approach is shown in Fig.~\ref{plot2}.
Our bound is very close to the one presented in Ref.~\cite{KARMEN2000}.

\end{appendix}

\end{document}